\documentclass[aps,amssymb,showpacs,twocolumn]{revtex4}
\usepackage{amssymb}
\usepackage{graphicx}
\usepackage{color}
\usepackage{textcomp}
\usepackage{ulem}
\begin{document}

\title{From Inverse to Delayed Magnetic Catalysis in Strong Magnetic Field }
\author{Shijun Mao}
\affiliation{School of Science, Xian Jiaotong University, Xian 710049, China}

\begin{abstract}
We study magnetic field effect on chiral phase transition in a Nambu--Jona-Lasinio model. In comparison with mean field approximation containing quarks only, including mesons as quantum fluctuations in the model leads to a transition from inverse to delayed magnetic catalysis at finite temperature and delays the transition at finite baryon chemical potential. The location of the critical end point depends on the the magnetic field non-monotonously.
\end{abstract}

\date{\today}
\pacs{21.65.Qr, 25.27.Nq, 75.30.Kz, 11.30.Rd}
\maketitle

The research on Quantum Chromodynamics (QCD) phase structure at finite temperature is recently extended to including external magnetic field, due to its close relation to high energy nuclear collisions and cosmological phase transitions. From lattice simulations of QCD at magnetic field $eB<1$ GeV$^2\sim 55\ m_\pi^2$~\cite{la1,la2,la3,la4}, while the chiral condensate is enhanced at low temperature which is called magnetic catalysis (MC), it is reduced at high temperature which leads to a decreasing critical temperature of chiral phase transition, named as inverse magnetic catalysis (IMC). Many scenarios~\cite{mc1,mc2,mc3,rev1,mu1,rev3,rev4,fukushima,kamikado,bf1,bf2,bf3,bf4,bf5,bf6,bf7,bf8,bf9,dmc2,dmc3,bf10,bf11,maocb,dmc1,mitb} are proposed to understand the MC and IMC effects. A straightforward question is how the chiral condensate behaves when the magnetic field increases further? Different hypotheses are recently introduced to study the chiral phase transition at extremely strong magnetic field. In some calculations the critical temperature turns around and increases when the magnetic field is sufficiently strong~\cite{bf2,bf9,dmc2,dmc3,dmc1}, which is named as delayed magnetic catalysis (DMC)~\cite{dmc2}. However, it is also argued that the critical temperature will keep decreasing~\cite{fukushima,bf3,mitb} which is supported by recent lattice simulation~\cite{lqcdb} where the IMC effect prevails up to $eB=3.25$ GeV$^2 \sim 180\ m_\pi^2$.

The magnetic field effect on QCD phase transitions at finite baryon chemical potential plays an important role in understanding the inner structure of compact stars. Because of the notorious sign problem, there is not yet precise result from lattice simulations. At zero temperature, the critical baryon chemical potential is argued to show a non-monotonous dependence on the magnetic field, and the MC effect is dominant only at sufficiently strong magnetic field, see review~\cite{mu1} and the references therein.

In this paper, we investigate the magnetic field effect on chiral phase transition in a Nambu--Jona-Lasinio model (NJL) beyond mean field approximation. We follow the theoretical framework developed in Ref.\cite{maocb} and focus on the DMC effect at finite temperature and baryon chemical potential. In NJL models, at finite temperature, the MC is described at mean field level~\cite{rev1,mu1,rev3,rev4}, and the IMC can be realized by including meson contribution to the quark self-energy~\cite{fukushima,maocb}. Since the DMC is predicted at extremely strong magnetic field, a convincing study in a non-renormalizable model with contact interactions depends strongly on the regularization scheme. We will take a covariant Pauli-Villars regularization which formally allows us to do momentum integrations in the whole region. We will focus on the magnetic field effect at $eB<\Lambda^2$ where $\Lambda$ is the cutoff introduced in the regularization scheme. Beyond this region the model is probably not applicable.

The SU(2) NJL model is defined through the Lagrangian density~\cite{njl1,njl2,njl3,njl4,njl5}
\begin{equation}
{\cal L}=\bar{\psi}\left(i\gamma_{\nu} D^{\nu}-m_0\right)\psi + \frac{G}{2}\left[\left( \bar{\psi} \psi \right)^2 + \left( \bar{\psi} i \gamma_5 {\vec \tau} \psi \right)^2  \right],
\end{equation}
where the covariant derivative $D^{\nu}=\partial^\nu+i Q A^\nu$ couples quarks to the external magnetic field ${\bf B}=(0, 0, B)$ along the $z$-axis, $Q=diag(Q_u, Q_d)=diag(2e/3,-e/3)$ is the quark charge matrix in flavor space, and $G$ is the coupling constant in the scalar and pseudo-scalar channels. In chiral limit with vanishing current quark mass $m_0=0$, the SU(2)$_L\otimes$SU(2)$_R$ symmetry is broken down to U(1)$_L\otimes$U(1)$_R$ by the magnetic field ${\bf B}$, and the number of Goldstone modes is reduced from 3 to 1. In the chiral symmetry breaking phase, quarks obtain mass $m=m_0-G \langle \bar q q\rangle$ from the chiral condensate $\langle \bar q q \rangle$. The mean field approximation for quarks together with the random phase approximation for mesons can describe well the chiral thermodynamics of hot and dense quark-meson plasma~\cite{zhuang}.

Describing quarks at mean field level and including mesons which are quantum fluctuations above the mean field in the model, the thermodynamic potential of the quark-meson plasma can be generally written as
\begin{equation}
\Omega={m^2\over 2G}+\Omega_q+\sum_M\Omega_M,
\end{equation}
The three terms are respectively the contributions from the condensates, quarks and mesons (isospin singlet $\sigma$ and triplet $\pi_0$ and $\pi_\pm$). The quark mass $m$ is determined by minimizing the thermodynamic potential $\partial\Omega/\partial m=0$, which leads to the gap equation,
\begin{equation}
m\left( \frac{1}{2G}+\frac{\partial \Omega_q}{\partial m^2}+\sum_M\frac{\partial \Omega_M}{\partial m^2}\right)=0
\label{gap1}
\end{equation}
in chiral limit.

Obviously, the quark mass $m$ from the gap equation(\ref{gap1}) is different from the mean field one $m_{mf}$ determined by $m_{mf}( 1/(2G)+\partial\Omega_q/\partial m_{mf}^2)=0$. Supposing the fluctuations induced correction is small, $|m-m_{mf}|/m_{mf}<<1$, we can expand the meson thermodynamics in terms of the correction~\cite{zhuang},
\begin{equation}
\Omega_M=\sum_n{\frac{1}{n!} \frac{\partial^n \Omega_M}{\partial (m^2)^n}\Big|_{m_{mf}^2}\left(m^2-m_{mf}^2\right)^n}.
\end{equation}
Keeping only the first two terms with $n=0,1$, the gap equation takes the same form as the mean field one,
\begin{equation}
\label{gap}
m\left( \frac{1}{2G'}+\frac{\partial \Omega_q}{\partial m^2}\right) =0,
\end{equation}
and the meson contribution is reflected only in an effective coupling constant $G'$,
\begin{equation}
\frac{1}{2G'} =\frac{1}{2G}+\sum_M\frac{\partial \Omega_M}{\partial m^2}\Big|_{m^2_{mf}}.
\end{equation}
Different from the original coupling $G$ which is a constant, the effective coupling $G'$ is a function of temperature, chemical potential and magnetic field. Under this approximation, the quark-meson system is treated effectively as a quark system.

We now consider the detailed expression for the quark and meson thermodynamics. This is done in Ref.\cite{maocb} at finite temperature $T$ and external magnetic field $B$. Including the baryon chemical potential $\mu_B$, the quark part at mean field level reads
\begin{eqnarray}
\Omega_q &=& -3 \sum_{q=u,d}\sum_n \alpha_n \int \frac{d p_z}{2\pi} \frac{|Q_q B|}{2\pi}\bigg[\frac{E_q^++E_q^-}{2}\nonumber\\
 &&+ T \ln \left(\left(1+e^{-E_q^+/T}\right)\left(1+e^{-E_q^-/T}\right)\right)\bigg]
\end{eqnarray}
with the spin degeneracy factor $\alpha_n=2-\delta_{n,0}$ and dispersion relations $ E^\pm_q=\sqrt{p^2_z+2 n |Q_q B|+m^2}\pm \mu_B/3$.  Under random phase approximation to construct mesons and pole approximation to take mesons as quasi-particles~\cite{zhuang}, the meson thermal contribution can be simply written as
\begin{equation}
\Omega_M = \int \frac{d^3 {\bf k}}{(2\pi)^3} \left(\frac{E_M}{2} +T \ln(1-e^{-E_M/T}) \right)
\end{equation}
with meson energy $E_M = \sqrt{m_M^2+k_3^2+v^2_\perp (k_1^2+k_2^2)}$, where the meson mass $m_M$ and quark-meson coupling constant $g_{q\bar q M}$ in the transverse velocity $v^2_\perp =\left(g^0_{q {\bar q} M}\right)^2/\left(g^1_{q {\bar q} M}\right)^2$ are determined by~\cite{njl2,njl3,njl4,njl5,zhuang,ritus6,ritus5}
\begin{eqnarray}
\label{pole}
&& 1-G\Pi_M(k_0^2=m^2_M,{\bf k}^2=0)=0,\nonumber\\
&& \left(g^\mu_{q\bar q M}\right)^2 =\left[g^{\mu\mu} \frac{d \Pi_M (k_0^2,{\bf k}^2)}{d k^2_\mu}\bigg|_{k^2=(m^2_M, 0)}\right]^{-1}
\end{eqnarray}
with $g^{\mu\nu} = diag(1,-1,-1,-1)$. The detailed expression for the meson polarization function (quark bubble) $\Pi_M(k_0^2,{\bf k}^2)$ at finite magnetic field can be found in Ref.\cite{ritus6,maocb}. Due to the introduction of the external magnetic field, the space is no longer isotropic, and the meson velocity depends on the moving direction.

Considering that the external magnetic field leads to discrete Landau levels and anisotropy in momentum space, we take a covariant Pauli-Villars regularization scheme, where the quark momentum runs formally to infinity and the divergence is removed by cancelation among the subtraction terms~\cite{njl2,njl3,njl4,njl5}. Under this regularization scheme, the law of causality is guaranteed and the MC effect at low temperature and IMC effect at high temperature are obtained in the magnetic field region $20 m_\pi^2 < eB < 50 m_\pi^2$~\cite{maocb}. The two parameters, coupling constant $G$=9.94 GeV$^{-2}$ and cutoff $\Lambda=$ 1127 MeV, in chiral limit are determined by fitting the chiral condensate $\langle \bar q q \rangle=-(250 \text{MeV})^3$ and pion decay constant $f_\pi=93$ MeV in vacuum. In this case, the reasonable magnetic field region in the NJL model is constrained by the condition $eB < \Lambda^2 \simeq 70\ m_\pi^2$.

We focus on the chiral breaking phase at low temperature and chemical potential where the meson degrees of freedom play a dominant role. Note that, applying random phase approximation to construct mesons, the quark propagator in quark bubbles is at mean field level~\cite{njl2,njl3,njl4,njl5}. By comparing the mean field gap equation for quark mass $m_{mf}$ with the pole equation (\ref{pole}) for neutral mesons $\sigma$ and $\pi_0$, we have the simple relations in the chiral symmetry breaking phase $m_{\pi_0} = 0$ and $m_\sigma = 2m_{mf}$. This indicates that $\pi_0$ is the Goldstone mode corresponding to the spontaneous chiral symmetry breaking and $\sigma$ is massive. As for charged pions $\pi_\pm$, the calculation for their polarization functions in an external magnetic field is much more complicated than that for neutral mesons $\pi_0$ and $\sigma$. However, they are no longer the Goldstone modes at nonzero magnetic field and become massive, especially at strong magnetic field. Therefore, for the discussion of DMC which is expected to happen at $eB > 50\ m_\pi^2$, we can safely neglect the contribution of charged pions and $\sigma$ to the thermodynamics of the system in the chiral breaking phase, and take into account the Goldstone mode $\pi_0$ only.
\begin{figure}[hb]
\centering
\includegraphics[width=7.5cm]{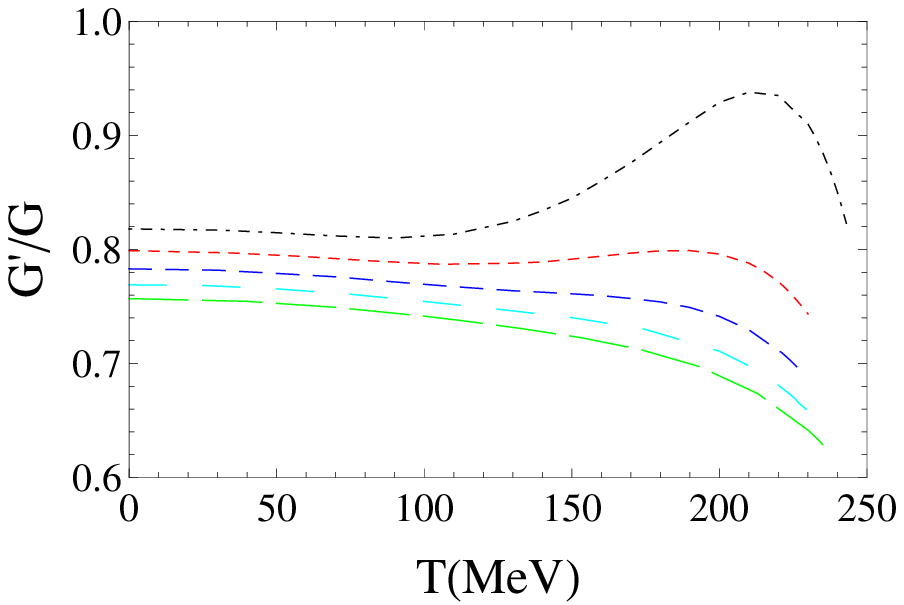}
\includegraphics[width=7.5cm]{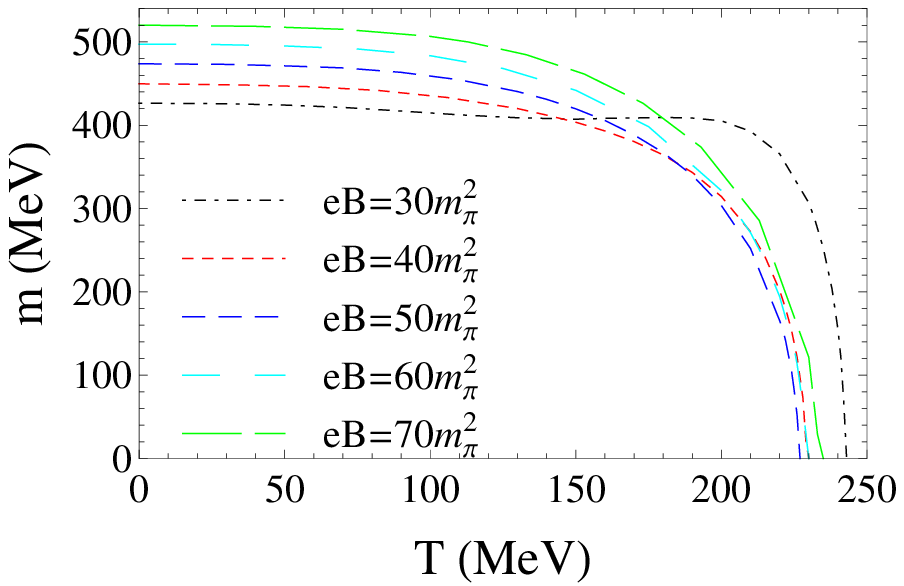}
\caption{ (Color online) The temperature dependence of the scaled effective coupling $G'/G$ (upper panel) and quark mass $m$ (lower panel) at zero baryon chemical potential and different magnetic field. }
\label{fig1}
\end{figure}
\begin{figure}[hbt]
\centering
\includegraphics[width=7.5cm]{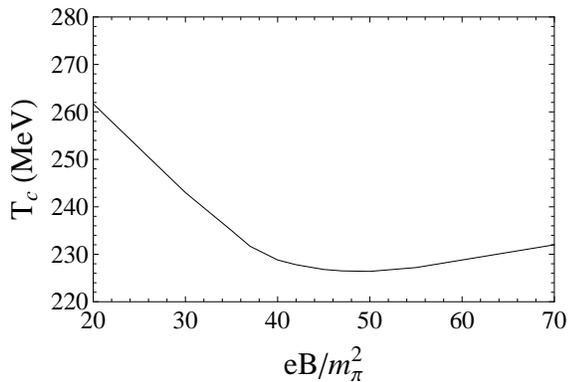}
\caption{ The critical temperature $T_c$ of chiral symmetry restoration as a function of magnetic field at zero baryon chemical potential. }
\label{fig2}
\end{figure}

The magnetic field effect at mean field level is controlled by the gap equation with a constant coupling $G$, which leads to the MC effect on quarks and the strength of MC effect is monotonously enhanced with increasing magnetic field~\cite{rev1,mu1,rev3,rev4}. 
Including meson contribution which is reflected in the medium dependent coupling $G'(T,\mu_B,B)$, the meson dressed quark mass is determined by the new gap equation(\ref{gap}), with parameters $T,\ \mu_B,\ B,\ G'$. Given $T,\ \mu_B, \ B$, we observe $G'/G<1$, the magnetic inhibition (MI) effect of mesons~\cite{fukushima}, seen in Fig.\ref{fig1} upper panel. The magnetic field dependence of critical temperature $T_c(B)$ comes from the competition between MC effect of quarks and MI effect of mesons. At $20 m_\pi^2 < eB < 50 m_\pi^2$ the magnetic inhibition of mesons is dominant, the result of the competition is the IMC phenomena, $T_c$ decreasing with increasing $B$~\cite{maocb}. However, from the temperature behavior of the effective coupling in the magnetic field region $30 m_\pi^2 < eB < 70 m_\pi^2$ shown in the upper panel of Fig.\ref{fig1}, while it keeps decreasing with increasing magnetic field at any temperature, the change becomes slower with stronger magnetic field $B$, especially around the critical temperature $T_c$. This indicates a saturated magnetic inhibition effect of mesons at extremely strong magnetic field. In this case, the monotonously strengthened MC effect on quarks dominates the competition and may lead to the MC phenomena, that is $T_c$ increases with increasing $B$ at sufficiently strong magnetic field. This is confirmed by the magnetic field dependence of the meson dressed quark mass $m$ shown in the lower panel of Fig.\ref{fig1}. At low temperature, the quark mass is always enhanced by external magnetic field, due to the stronger magnetic catalysis effect on quarks than magnetic inhibition of mesons. Around the critical temperature, however, the quark mass drops down first and then goes up with increasing magnetic field. The reason is the rapidly decreasing $G'$ due to strong magnetic inhibition of mesons at $eB < 50 m_\pi^2$ and the slowly changed $G'$ due to the saturated magnetic inhibition of mesons at $eB > 50 m_\pi^2$. The turning point is located around $eB \simeq 50 m_\pi^2$.
\begin{figure}[hb]
\centering
\includegraphics[width=7.5cm]{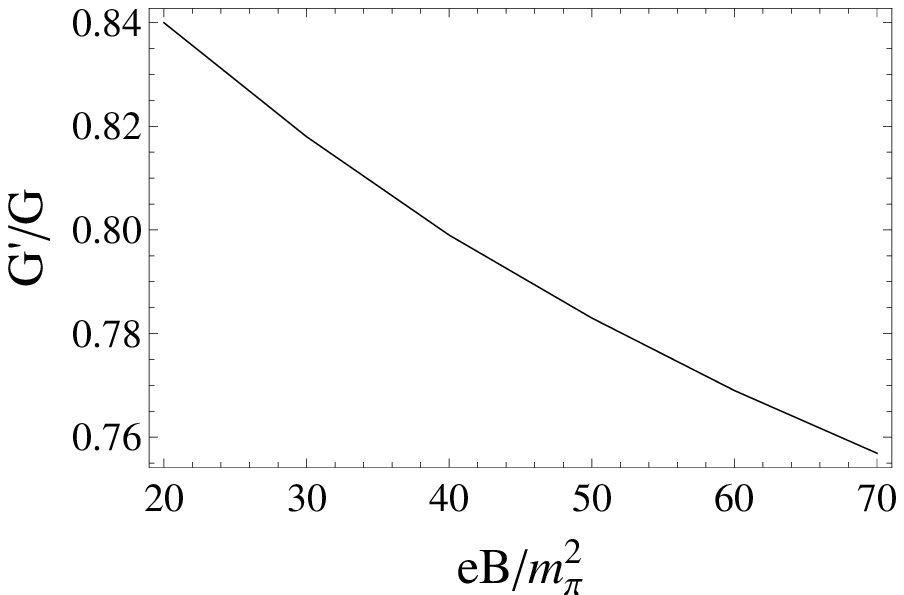}
\includegraphics[width=7.5cm]{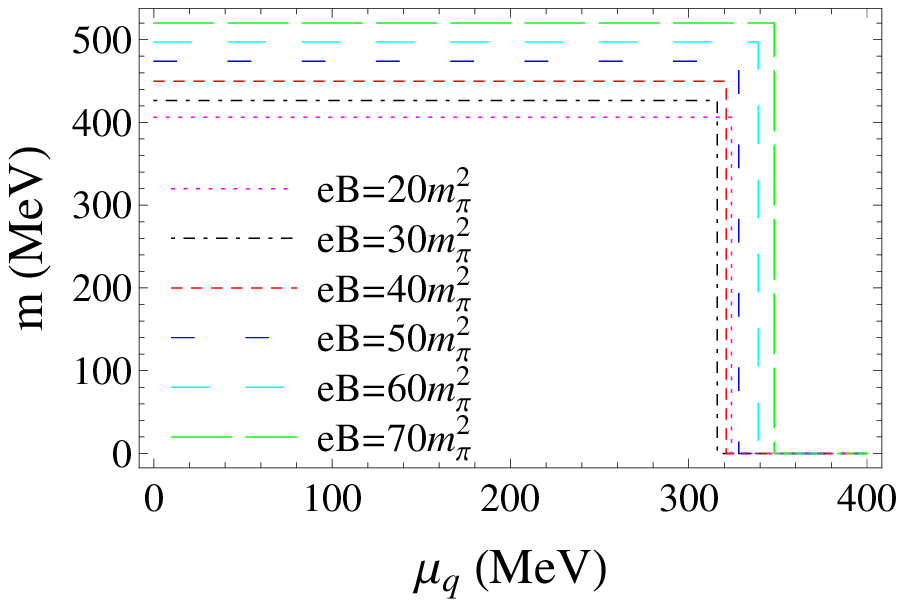}
\caption{(Color online) The scaled effective coupling $G'/G$ (upper panel) as a function of magnetic field at zero temperature, and quark mass $m$ (lower panel) as a function of quark chemical potential $\mu_q=\mu_B/3$ at zero temperature and different magnetic field. }
\label{fig3}
\end{figure}
\begin{figure}[hbt]
\centering
\includegraphics[width=7.5cm]{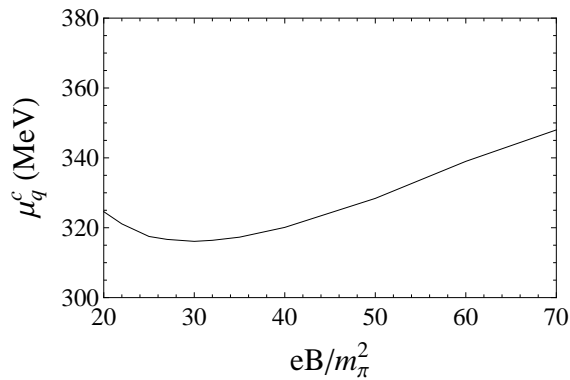}
\caption{ The critical quark chemical potential $\mu_q^c=\mu_B^c/3$ of chiral symmetry restoration as a function of magnetic field at zero temperature. }
\label{fig5}
\end{figure}
\begin{figure}[hbt]
\centering
\includegraphics[width=7.5cm]{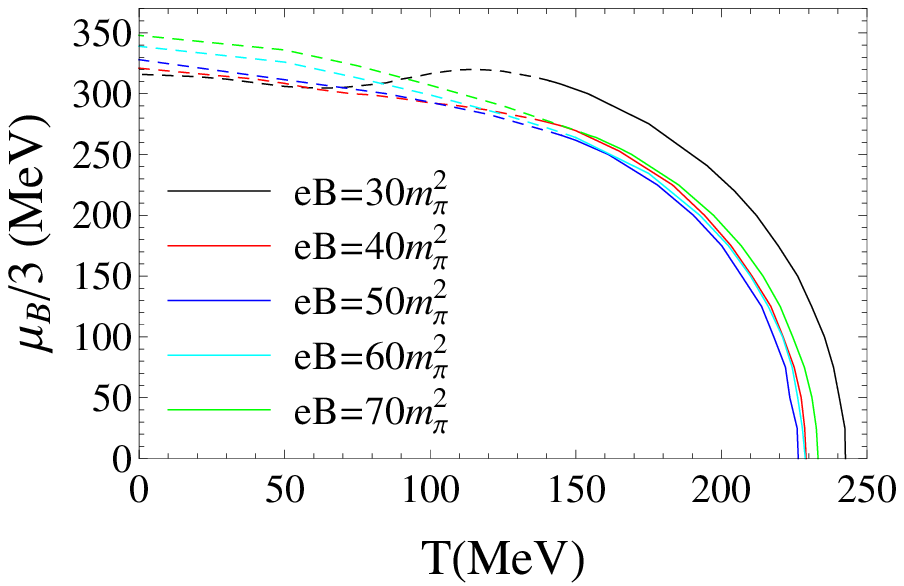}
\includegraphics[width=7.5cm]{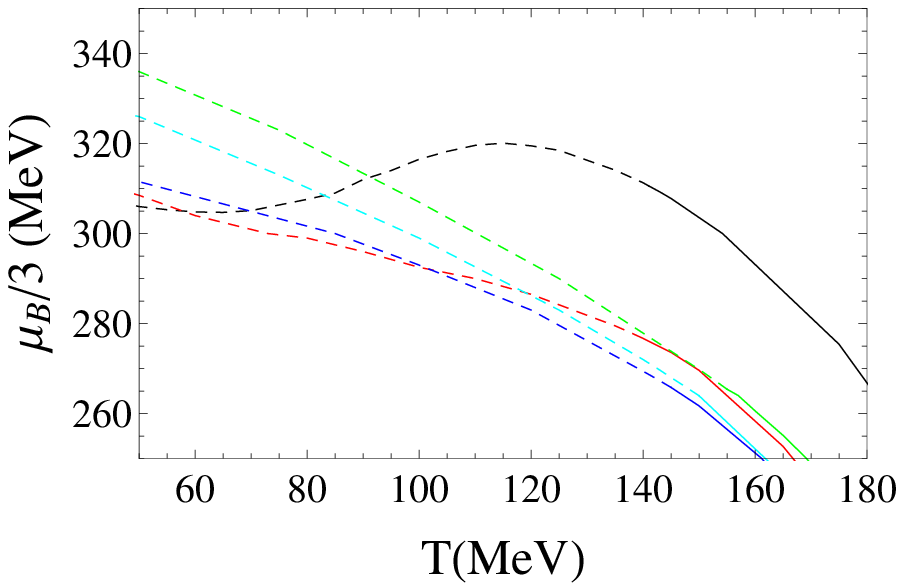}
\caption{ (Color online) The chiral phase transition lines in $T$-$\mu_B$ plane at different magnetic field. The solid and dashed lines indicate respectively first and second order phase transitions. Lower panel is the phase diagram near the critical end point.  }
\label{fig6}
\end{figure}

To expressly demonstrate the magnetic field effect on chiral phase transition, we plot the critical temperature $T_c$, defined by $m(T_c,\mu_B=0,B)=0$ in chiral limit, as a function of the scaled magnetic field $eB/m_\pi^2$ in Fig.\ref{fig2}. It decreases at $eB < 50 m_\pi^2$ and then turns around at about $eB = 50 m_\pi^2$ and increases at $eB > 50 m_\pi^2$. This shows clearly a transition from IMC to DMC with increasing magnetic field and is consistent with the results from other model calculation~\cite{bf9,dmc1}, Dyson-Schwinger equation~\cite{dmc2}, functional renormalization group~\cite{bf2} and two-color lattice QCD~\cite{dmc3}. However, it is different from the conclusion of $1+1+1$-flavor lattice QCD~\cite{lqcdb}. Note that, with nonzero magnetic field, the number of Goldstone modes is reduced from 3 to 1. This is probably the reason why the critical temperature here ($>$ 220 MeV) is higher than the typical value at vanishing magnetic field ($\simeq$ 170 MeV). We have numerically checked that, we can reduce the critical temperature by tuning the model parameters but it does not change qualitatively the transition from IMC to DMC with increasing magnetic field.

We now turn to the discussion at finite baryon chemical potential. At vanishing magnetic field, it is well known that the chiral symmetry restoration switches from a continuous phase transition at finite temperature to a first-order phase transition at finite baryon chemical potential: The order parameter, namely the quark mass, suddenly jumps down to zero at the critical chemical potential~\cite{njl2,njl3,njl4,njl5,zhuang}. At nonzero magnetic field, while this sudden jump of the order parameter persists, there exists the change from IMC to DMC with increasing magnetic field even in mean field approximation~\cite{mu1}. Going beyond the mean field by including meson contribution, the mesons do not carry baryon charge, and their chemical potential dependence comes from their constituents. At zero temperature, the mean field quark mass keeps as a constant in the whole chiral symmetry breaking phase, and therefore, the effective coupling $G'$ does not depend on $\mu_B$ and is a function of the magnetic field only, shown in Fig.\ref{fig3} upper panel. It is always less than the original coupling $G$, $G'/G<1$, and decreases continuously. With the known effective coupling, we resolve the gap equation (\ref{gap}) beyond mean field, and the meson dressed quark mass is plotted in Fig.\ref{fig3} lower panel as a function of the quark chemical potential $\mu_q=\mu_B/3$ at $T=0$ and different magnetic field. The dressed quark mass behaves similar to the mean field case, keeping as a constant first and then dropping down to zero suddenly. 
Compared with mean field calculation, we obtain the extra magnetic inhibition effect from mesons. Therefore, including mesons at finite baryon chemical potential delays the transition from IMC to DMC. This is clearly shown in the magnetic field dependence of the critical quark chemical potential $\mu_q^c=\mu_B^c/3$ defined as $m(T=0,\mu_q^c,B)=0$, shown in Fig.\ref{fig5}. It drops down first and then goes up continuously with increasing magnetic field. The turning point from IMC to DMC is located around $eB \simeq 30 m_\pi^2$, larger than the mean field value $\simeq 10 m_\pi^2$.

The calculation can be extended to finite temperature and chemical potential. The chiral phase transition lines in $T$-$\mu_B$ plane at different magnetic field are shown in Fig.\ref{fig6}. The transition from IMC to DMC at finite temperature or chemical potential happens at different magnetic field, and hence there are crossings in the phase transition line at different $B$. 
The location of the critical end point which links the continuous phase transition (dashed line) at high temperature and the first-order phase transition (solid line) at high chemical potential varies non-monotonously with magnetic field.

In summary, magnetic field in quark and meson part leads to magnetic catalysis and magnetic inhibition effect, respectively, and the competition between them controls the chiral phase transition. In mean field approximation with quarks only, there is always a magnetic catalysis at finite temperature, but there exists a transition from inverse magnetic catalysis to delayed magnetic catalysis at finite baryon chemical potential. Including mesons as quantum fluctuations above the mean field, the effective coupling among quarks becomes weaker and depends on the medium parameters (temperature, chemical potential and magnetic field). At finite temperature, the competition between the magnetic catalysis effect on quarks and magnetic inhibition of mesons leads also to the transition from IMC to DMC with increasing magnetic field. At finite baryon chemical potential, including mesons delays the transition and the turning point is located at a larger magnetic field than mean field situation. At finite temperature and chemical potential, the location of the critical end point of chiral phase transition depends non-monotonously on the magnetic field.

\noindent {\bf Acknowledgement:} The work is supported by the NSFC and China Postdoctoral Science Foundation Grants 11405122, 11575093 and 2014M550483.

\end{document}